\documentclass[twocolumn,tighten,astrosymb,trackchanges]{aastex631}

\newcommand{\UCSC}{\affiliation{Department of Astronomy and Astrophysics, University of California, Santa Cruz, CA 95064-1077, USA}}

\usepackage{CJK}
\usepackage{array}
\usepackage{multirow}

\newcommand\um{$\mu$m }
\newcommand\umm{$\mu$m}
\newcommand\grb{GRB~230307A }
\newcommand\grbb{GRB~230307A}
\newcommand\irkilo{AT~2023vfi}

\let\tablenum\relax
\usepackage{siunitx}
\usepackage{appendix}
\usepackage{amsmath,amsthm,amsfonts,amssymb,amscd}
\usepackage{booktabs}
\usepackage{tablefootnote}
\usepackage{apjfonts}
\usepackage{makecell}
\usepackage{graphicx}

\newcolumntype{P}[1]{>{\centering\arraybackslash}p{#1}}  

\def\msun{\hbox{\,$M_{\odot}$}}

\def\cm{\mbox{\,cm}}
\def\cm3{\mbox{\,cm$^{-3}$}}

\shorttitle{GRB~230307A Dust}
\shortauthors{Arunachalam, Macias, \& Foley}

\begin{document}

\title{GRB 230307A Formed No Dust or Was Not a Binary Neutron Star Merger}

\correspondingauthor{Prasiddha Arunachalam}
\email{parunach@ucsc.edu}

\author[0000-0002-6688-3307]{Prasiddha~Arunachalam}
\UCSC
\author[0000-0002-9946-4635]{Phillip~Macias}
\UCSC
\author[0000-0001-6706-2749]{Ryan~J.~Foley}
\UCSC

\begin{abstract}
We present a new analysis of the \textit{JWST} infrared spectra of \grbb~(\irkilo), a long gamma-ray burst (GRB) with an infrared excess and spectral lines suggestive of significant  heavy $r$-process production. The spectra, taken 29 and 61~days after the GRB trigger, have blackbody-like continua with $T_{\rm eff} \approx 550$~K and an emission line near $2.1$ \umm, previously attributed to [\ion{Te}{3}]. This line identification has been used as evidence for an $r$-process-powered kilonova (KN), despite no KN model producing a blackbody-like spectrum at late times. Such an infrared continuum could be emitted by newly formed dust, and we model the thermal emission to infer dust properties, including composition and mass.  We find that the emission requires at least 3--$6 \times 10^{-3}$~M$_{\sun}$ of carbon or silicate dust, which is inconsistent with $r$-process yields expected from a neutron star merger. Alternatively, the continuum could be from $2\times 10^{-3}$~M$_{\sun}$ of metallic iron dust, which is mildly consistent (at 3$\sigma$) with KN models, but such dust is unlikely to form in the expanding ejecta.  GRB~230307A's low late-time luminosity also constrains the amount of radioactive $^{56}$Ni produced to $<${}$2.6 \times 10^{-3}$~M$_{\sun}$ (3$\sigma$).  No KN model can simultaneously form the necessary dust for the infrared continuum and heavy elements for the [\ion{Te}{3}] line.  We conclude that the blackbody continuum is not due to dust emission, or \grb did not originate from a binary compact-object merger.
\end{abstract}

\keywords{GRB, Kilonova, compact-object merger, $r$-process nucleosynthesis}

\section{Introduction \label{sec:intro}}
The nearly simultaneous observations of GW170817 and GRB~170817A \citep{Abbott2017:mma, Abbott2017:grb, Goldstein2017, Savchenko2017} confirm that compact-object mergers can create short-duration gamma-ray bursts \citep{Paczynski1986,Eichler1989,Narayan1992}, supporting decades of theory and indirect observational evidence \citep[for review see][]{Nakar2007, Berger2014}.  The discovery of an associated optical/infrared (IR) kilonova (KN) counterpart, SSS17a or AT~2017gfo \citep{Coulter2017}, and its associated spectral and color evolution strongly suggest that heavy elements associated with the ``hard'' or ``strong'' $r$ process were produced in abundance \citep{Cowperthwaite2017, Drout2017, Kasen2017, Kasliwal2017, Pian2017, Smartt2017, Tanvir2017}.  These observations, in total, show that inspiraling and merging neutron stars emit gravitational waves (GWs) and can produce heavy $r$-process-enriched ejecta and relativistic jets \citep{Kilpatrick2017, Lazzati2018}.

Since the discovery of GW170817 and GRB~170817A/ AT~2017gfo, no multi-messenger GW event has been discovered.  Nevertheless, our understanding of $r$-process enrichment through compact-binary mergers has been improved by searching for KNe associated with GRBs \citep[e.g.,][]{Lamb2019, Rastinejad2022, 
Troja2022, Levan2024, Yang2024}  Since KNe have a low luminosity that peaks while an on-axis afterglow is still bright, one can only observe KNe for the closest GRBs with relatively low-luminosity afterglows.

\grb stands out as one of the brightest GRBs ever recorded, with a prompt fluence of $\sim3\times 10^{-3}$~erg~cm$^{-2}$ measured by Fermi-GBM (cite) and a duration of T$_{90} \approx 35$~s \citep{Dalessi2025}, classifying it observationally as a “long” GRB. Its sky position is $\sim$40~kpc from the center of a presumed host galaxy, a nearby spiral galaxy at $z = 0.0646$ \citep[a closer galaxy at $z = 3.87$ is excluded because of the lack of a Lyman break in the afterglow]{Levan2024, Yang2024}.  The afterglow observations in X-ray, optical, and radio reveal a surprisingly faint afterglow relative to its prompt fluence \citep{Levan2024, Yang2024}, somewhat resembling those of short GRBs with extended emission. A strikingly similar behavior was observed in GRB~211211A, which despite its long prompt duration ($T_{90} \approx 40$~s) showed a faint afterglow and KN-like emission \citep{Rastinejad2022,Troja2022,Yang2022}. This parallel suggests that some merger-driven GRBs may appear as long bursts. 

Late-time observations of \grbb’s ``thermal'' counterpart, \irkilo, provided the first opportunity for deep \textit{JWST} spectroscopy of a GRB. \textit{JWST} and ground-based follow-up at $\sim$29 and $\sim$61~days post-burst revealed a prominent IR excess, dominated by a red thermal continuum that extends beyond 4.5~$\mu$m \citep{Levan2024, Gillanders2023}. Spectra also show emission-like features near $\sim$2.2~$\mu$m, attributed to Te---produced at the top of the second peak of $r$-process nucleosynthesis \citep{Gillanders2023, Levan2024, Yang2024, Gill&Smartt2025}. At the presumed distance of its host, the transient’s light curve, temperature, and radius closely match the evolution seen in AT~2017gfo \citep{Yang2024}. These findings, along with the proposed detection of Te, a signature of $r$-process nucleosynthesis, have been interpreted as strong evidence that \grb originated from a compact-object merger accompanied by a KN.

KNe are not expected to be optically thick at the spectroscopic epochs of \grb, and no KN model predicts a pseudo-blackbody continuum at late times \citep[e.g.,][]{Hotokezaka2021,
Hotokezaka2022, Banerjee2025, Pognan2025}.  A potential explanation for the IR emission could be re-emission from hot dust \citep[see discussions in][]{Gillanders2023, Levan2024}. \citet{Takami2014} considered dust formation in KN ejecta and found that dust species composed of light elements such as carbon (and other lighter than iron) could potentially condense if those elements are sufficiently abundant; in contrast, they found that dust made purely of heavy $r$-process nuclei (lanthanides/actinides) is very hard to form because low number densities in the expanding ejecta. If dust exists close to the KN, it could absorb the ejecta’s emission and re-emit as a thermal IR continuum. The expected observational signature of a dust-powered transient is a smooth, featureless IR spectrum, similar to that seen for \irkilo.

\citet{Gall2017} modeled the dust extinction scenario for AT~2017gfo and identified stringent conditions for dust to play a dominant role in the electromagnetic emission from compact-object mergers. They found that reproducing the observed blue-to-IR reddening would require approximately $10^{-5}$ M$\odot$ of carbon dust to form within just a few days after the merger—far exceeding the $\sim$$10^{-9}$ M$\odot$ they obtained from kilonova wind models \citep{Rosswog2017}. In addition, the dust would need to maintain extremely high temperatures ($\sim$2000 K at 4 days, cooling to 1500 K after a week) to account for the IR luminosity. Given these discrepancies in dust mass and formation timescales, they concluded that $r$-process opacity from lanthanide-rich ejecta offer a more plausible explanation for the IR emission in GW170817 than newly formed dust.

In addition to dust newly formed in the KN ejecta, IR emission in GRBs can also originate from pre-existing dust heated by the GRB’s prompt or afterglow radiation. A similar mechanism may also be occur for \grbb, as \citet{Waxman2025} argued in the case of GRB~211211A (with similar high energy characteristics to \grbb), that the observed IR excess arose not from a a KN, but from pre-existing ambient dust-such as in a molecular cloud- heated by the GRB.

In this paper, we examine the possibility of the IR continuum in the spectra of \grb originating from dust emission.  In Section~\ref{sec:dust_model_fits}, we fit dust models to the IR emission to derive physical characteristics of the dust.  We compare these abundances to those produced by KN models in Section~\ref{sec:kilonova_model_comparison}, finding that no KN model can produce the necessary amount of light elements.  We discuss the implications of these findings in Section~\ref{sec:discussion} and conclude in Section~\ref{sec:conclusions}.

\section{Constraining Dust Properties for \irkilo}
\label{sec:dust_model_fits}

\begin{figure*}
    \centering
    \includegraphics[width=0.95\linewidth]{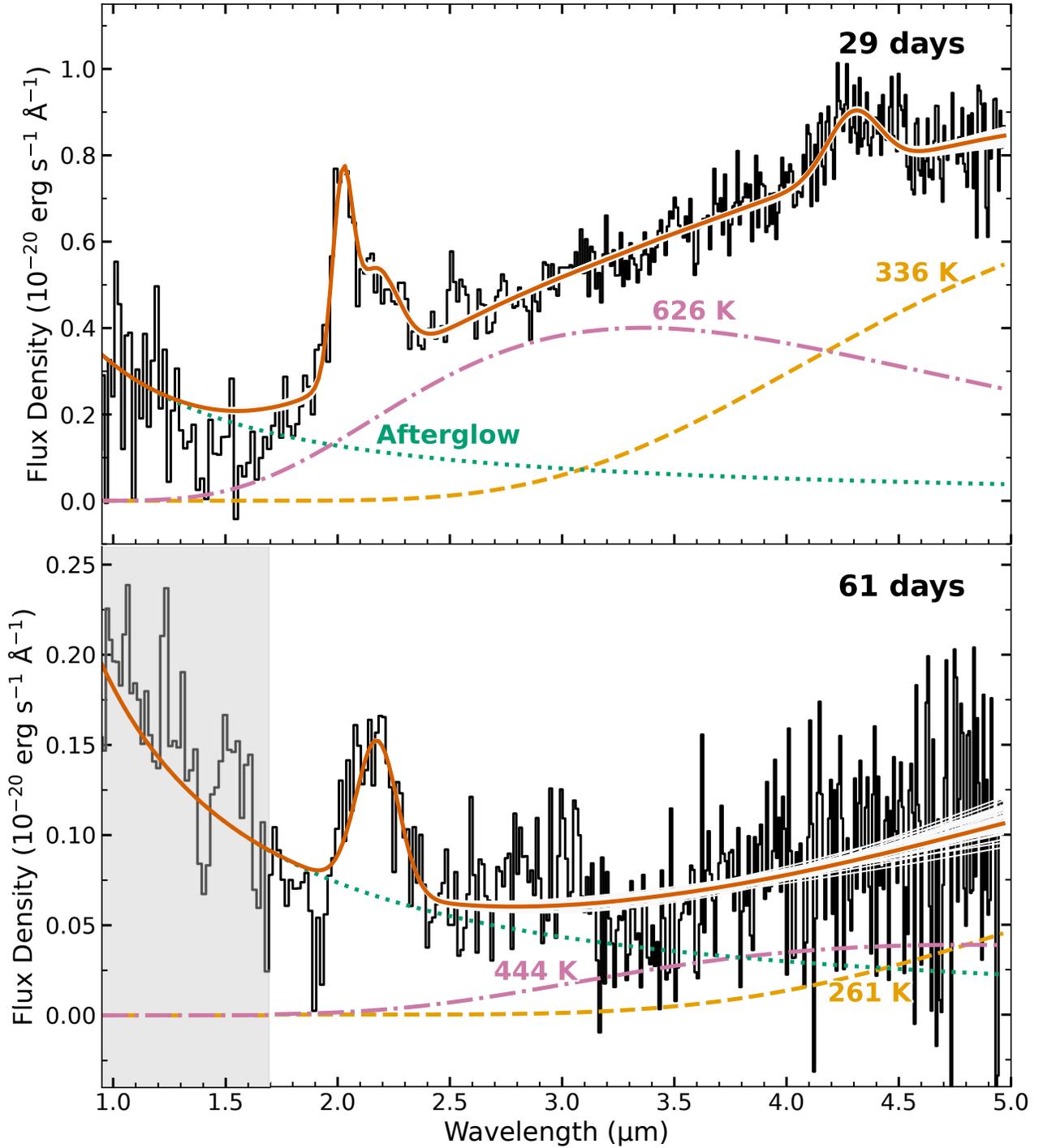}

    \caption{{\it JWST}/NIRSpec spectra of \irkilo{} taken 29 (top) and 61~days (bottom) after the GRB trigger, as reduced by 
    \citet[black curves]{Gill&Smartt2025}.  Overplotted in burnt red are the best-fit models for each spectrum.  Those spectra are a combination of a power-law component 
    (green dotted curve), two carbonaceous dust components (gold dashed and purple dot-dashed curves), and Gaussian profiles (not shown).  The temperatures of the dust components are listed on the plot: 336 and 626~K for the +29-day spectrum and 261 and 444~K 
    for the +61-day spectrum. The earlier spectrum has three Gaussian components, two centered near 2.1~\um{} and one centered near 4.2~\um.  The later spectrum has 
    only a single Gaussian component, centered near 2.1~\um{}.  While fitting the +61-day spectrum, the data in the grey region ($\lambda < 1.7$~\um) were not used.}
\label{fig:mcmcfit}
\end{figure*}

\textit{JWST} observed \irkilo\, +28.9 and +61.4~observer-frame days after the GRB trigger, obtaining IR photometry and spectra (GO--4434 and GO--4445; PI Levan).  These data were first presented by \citet{Levan2024} and \citet{Yang2024}. The NIRSpec spectra were contaminated by a nearby faint galaxy in the first epoch and a bright nearby star.   \citet{Gill&Smartt2025} re-reduced the data taking special care to mitigate these issues, producing spectra that have a higher signal-to-noise ratio, fewer artifacts, better calibration, and spectrophotometry consistent with the NIRCam imaging.  Despite these improvements, the earlier spectrum is still contaminated by light from the galaxy, as evidenced by narrow emission lines.  Although the +29-day spectrum may be contaminated by this galaxy, its allowed spectrum shape and brightness mean it cannot qualitatively affect any results.  In this analysis, we make use of these publicly available spectra, which are presented in Figure~\ref{fig:mcmcfit}.

Both spectra have several distinct components: a power-law (non-thermal) component that is strongest at shorter wavelengths and is consistent with coming from an afterglow, broad emission features centered near  $\sim$2.1~\um in both spectra and at $\sim$4.4~\um that is only visible in the +29-day spectrum, and a red, featureless continuum that peaks redward of the spectral coverage (i.e., $>$5~\umm) that is consistent with blackbody emission. The ``blackbody'' component is particularly prominent at the earlier epoch, where the continuum appears flat and faint at shorter wavelengths but rises steeply for $\lambda > 1.5$~\umm, consistent with cool thermal emission with  $T \approx 600$~K.  \citet{Gill&Smartt2025} constrain the afterglow component using the X-ray data \citep{Yang2024} extrapolated to IR wavelengths. They then fit a single blackbody spectrum to capture the thermal continuum at longer wavelengths. The emission features are modeled as the superposition of two Gaussian profiles for the $\sim$2.1~\um  feature at +29~days, and as single Gaussian profiles for the $\sim$4.4~\um feature at +29~days and the $\sim$2.1~\um feature at +61~days. \par

We perform a Markov-Chain Monte Carlo (MCMC) analysis to model the different components of the observed spectra. As a validation step, we first implement the model and assumptions described by \citet{Gill&Smartt2025}, which include the afterglow, blackbody continuum, and Gaussian emission components with wide prior ranges. To replicate the previous work, we fit the distance-corrected spectrum.  Using this setup, we successfully reproduce their results to within $1\sigma$ of the reported best-fit values, confirming the robustness of our MCMC implementation. \par

We then replace the blackbody continuum with a physically motivated dust spectral-energy distribution, and fit the data in flux space to incorporate the distance to \grb and its uncertainties. The distance has a Gaussian prior with a mean of $\mu_d = 8.99\times10^{26}~\mathrm{cm}$ and standard deviation $\sigma_d = 1.12\times10^{25}\ \mathrm{cm}$. The dust parameters (temperature and mass) use broad, non-negative uniform priors.  The afterglow parameters have flat priors based on the best-fit parameters from \citet{Gill&Smartt2025}, but widths set to twice their reported $1\sigma$ uncertainties. The emission features are modeled as Gaussians with uniform non-negative priors on their amplitudes, centroids, and line widths.

Assuming the blackbody-like emission seen in the spectrum is due to dust, we implement a modified blackbody component that represents the flux from dust emission \citep{Fox2010},\vspace{-0.2cm}
\begin{align}
     F_{\nu, d} &= \frac{M_d B_\nu(T_d) \kappa_\nu }{D^2} 
\end{align}
where $M_d$ is the dust mass, $B_\nu(T_d)$ is the Planck function at a dust temperature $T_d$, $D$ is the distance to the GRB, and $\kappa_\nu$  is the dust mass absorption coefficient, given by
\begin{align}
     \kappa_\nu &= \frac{3}{4 a \rho_d} Q_{\nu}(a).
\end{align}
Here, $Q_\nu$ is the emission efficiency as a function of grain radius $a$, and $\rho_d$ is the density of the dust grains.

Although a single-component dust model is the simplest choice, we find it is insufficient to correctly reproduce the observed spectra in the 2--3~\um range. Quantitatively, the two-component carbon dust model (16 parameters) yields a reduced $\chi^2$ ($\chi^2_\nu$) of 1.92, whereas the single-component model (14 parameters) has $\chi^2_\nu = 2.94$, strongly favoring the two-component model despite the modest increase in complexity. The corresponding values for silicate and iron dust compositions are similar, and likewise favor a two-component model. We therefore adopt the two-component dust model consisting of a single dust species at two distinct temperatures, which we refer to as ``warm'' and ``cold'' components. We test carbonaceous (graphite), silicate, and pure iron dust compositions, assuming a grain size of $0.1$~\umm.  For the carbonaceous and silicate dust, we use $Q_\nu$ and $\rho_d$ values from \citet{Draine1984}.  For the iron dust, we calculate $Q_\nu$ directly using the \citet{Ordal1988} optical constants and assuming spherical grains.  \citet{Kemper2002} examined non-spherical distributions of iron grains, but the chosen shape distribution is severely biased \citep{Draine2021}, and thus any examination of non-spherical grains would require new calculations by dust experts.  The results for all the dust-types are summarized in Table~\ref{tab:individual_fits}.

\begin{figure*}
    \centering
    \includegraphics[width=\linewidth]{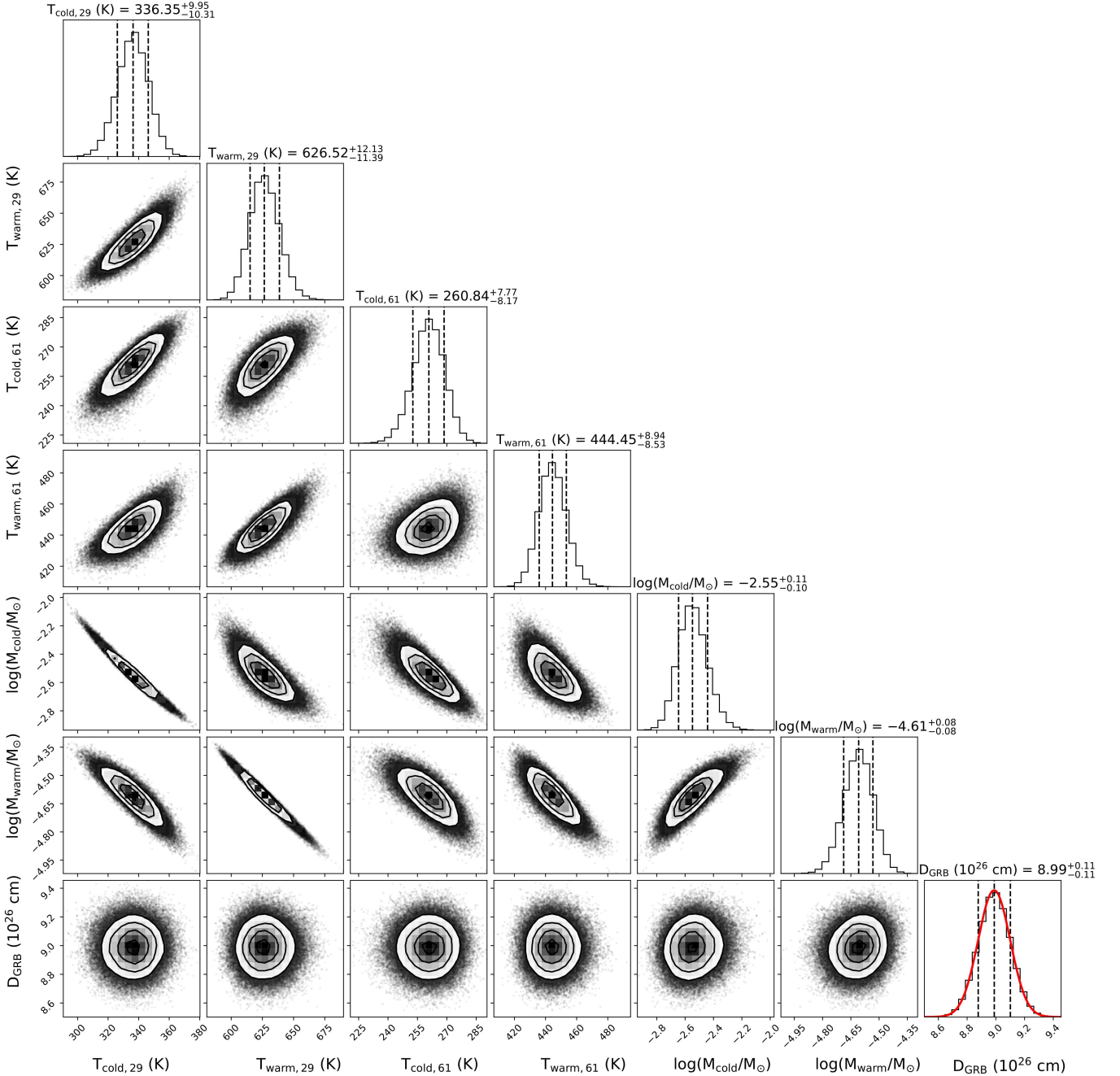}
\caption{MCMC corner plot showing the two-component dust model fits to the blackbody-like emission in \irkilo, using carbonaceous dust (grain size 0.1~\umm). The model fits the 29-day and 61-day data simultaneously, assuming fixed dust masses between the two epochs. This corner plot corresponds to the fit shown in Figure~\ref{fig:mcmcfit}, with the afterglow and emission features fixed to the values described in \citet{Gill&Smartt2025}.}
    \label{fig:corner}
\end{figure*}

For all models, we find that dust masses for the warm and cold components are consistent between the two epochs.  This implies that any dust creation or destruction is not detected between the two {\it JWST} epochs.  This result motivated fitting the data with dust masses in each component being identical in both epochs but allowing the dust temperature to vary between epochs.  Results for this model with all dust compositions are presented in Table~\ref{tab:joint_fits} and a corner plot for the carbonaceous dust model is shown in Figure~\ref{fig:corner}.  For the rest of the analysis, we focus on the results from this model, but the qualitative results are identical regardless of model choice.

For all dust types, the cold component dominates the overall mass budget. The cold ($T_{29}=336 \pm 10$~K, $T_{61}=261 \pm 8$~K) component of the carbonaceous (graphite) dust model contains $(2.8 \pm 0.7)\times10^{-3}$~M$_{\odot}$ of dust, a factor of $\sim$110 more than that of the warm component, which contains only $(2.5 \pm 0.4)\times10^{-5}$~M$_{\odot}$.  
For silicate dust, the cold component contains $(6.0 \pm 1.9)\times10^{-3}$~M$_{\odot}$, while the warm component contains $(8.4 \pm 1.4)\times10^{-5}$~M$_{\odot}$.  
The metallic Fe model yields $(1.9 \pm 0.4)\times10^{-3}$~M$_{\odot}$ in the cold component and $(1.6 \pm 0.3)\times10^{-5}$~M$_{\odot}$ in the warm component.

From the measured dust masses, we infer the mass fraction of the ejecta composed of individual elements (C, Si, Mg, O, and Fe), depending on the assumed dust composition.  
For this calculation, we adopt a total ejecta mass of $0.059 \pm 0.011~M_\odot$ \citep{Gillanders2023}.  
For carbonaceous (graphite) dust, the entire dust mass is attributed to carbon, yielding a mass fraction of $4.7 \pm 1.4$\% of the ejecta in C.  
For silicate dust, the derived mass fractions depend on the assumed composition---fosterite (Mg$_{2}$SiO$_{4}$), enstatite (MgSiO$_{3}$), or silica (SiO$_{2}$)---as listed in Table~\ref{tab:abundances}. 
Across these compositions, we find that Si, Mg, and O account for approximately $2.0$--$4.8$\%, $0$--$3.5$\%, and $2.8$--$5.0$\% of the ejecta mass, respectively.  
For metallic iron dust, the inferred Fe mass fraction is $3.2 \pm 0.9$\% of the total ejecta.

\begin{deluxetable}{c c c c c}\label{tab:individual_fits}
\tablecolumns{5}
\tablewidth{0pt}
\tabletypesize{\scriptsize}
\tablecaption{Dust Parameters for Each Spectroscopic Epoch.}
\tablehead{
\colhead{Epoch} & \colhead{Component} & \colhead{Temperature (K)} & \colhead{Dust Type} & \colhead{Dust Mass ($M_\odot$)}
}
\startdata
\multirow{6}{*}{29~days}
  & \multirow{3}{*}{Cold}
    & $356.8 \pm 10.4$  & Carbon (graphite) & $(2.0 \pm 0.4)\times10^{-3}$ \\
  & & $393.7 \pm 16.0$  & Silicate          & $(3.4 \pm 0.9)\times10^{-3}$ \\
  & & $376.6 \pm 12.5$  & Metallic iron     & $(1.3 \pm 0.3)\times10^{-3}$ \\
\cline{2-5}
  & \multirow{3}{*}{Warm}
    & $686.3 \pm 21.0$  & Carbon (graphite) & $(1.2 \pm 0.3)\times10^{-5}$ \\
  & & $781.2 \pm 27.7$  & Silicate          & $(4.2 \pm 1.1)\times10^{-5}$ \\
  & & $763.9 \pm 28.9$  & Metallic iron     & $(8.4 \pm 2.4)\times10^{-6}$ \\
\hline
\multirow{6}{*}{61~days}
  & \multirow{3}{*}{Cold}
    & $293.8 \pm 18.5$  & Carbon (graphite) & $(1.4 \pm 1.0)\times10^{-3}$ \\
  & & $300.6 \pm 25.8$  & Silicate          & $(4.2 \pm 3.7)\times10^{-3}$ \\
  & & $294.5 \pm 20.6$  & Metallic iron     & $(1.5 \pm 1.1)\times10^{-3}$ \\
\cline{2-5}
  & \multirow{3}{*}{Warm}
    & $780.5 \pm 27.6$  & Carbon (graphite) & $(4.1 \pm 1.1)\times10^{-7}$ \\
  & & $779.2 \pm 26.0$  & Silicate          & $(3.2 \pm 0.8)\times10^{-6}$ \\
  & & $777.4 \pm 26.0$  & Metallic iron     & $(5.9 \pm 1.4)\times10^{-7}$ \\
\enddata
\tablecomments{Temperatures and masses derived from two-component dust fits at 29 and 61 days. Uncertainties are $1\sigma$ errors from MCMC posteriors.}
\end{deluxetable}

\begin{deluxetable}{c c c c c}\label{tab:joint_fits}
\tablecolumns{5}
\tablewidth{0pt}
\tabletypesize{\scriptsize}
\tablecaption{Dust Parameters When Fixing the Dust Mass Across Both Spectroscopic Epochs.}
\tablehead{
\colhead{Type} & \colhead{Component} &
\colhead{$T_{29}$ (K)} &
\colhead{$T_{61}$ (K)} &
\colhead{$M_{\rm dust}$ ($M_\odot$)}
}
\startdata
\multirow{2}{*}{Carbon}   & Cold & $336.3 \pm 10.2$ & $260.9 \pm 8.1$  & $(2.8 \pm 0.7)\times10^{-3}$ \\
                          & Warm & $626.4 \pm 11.6$ & $444.4 \pm 8.6$  & $(2.5 \pm 0.4)\times10^{-5}$ \\
\hline
\multirow{2}{*}{Silicate} & Cold & $359.1 \pm 15.8$ & $271.3 \pm 12.0$ & $(6.0 \pm 1.9)\times10^{-3}$ \\
                          & Warm & $710.9 \pm 13.9$ & $482.3 \pm 9.5$  & $(8.4 \pm 1.4)\times10^{-5}$ \\
\hline
\multirow{2}{*}{Metallic Fe} & Cold & $354.9 \pm 11.3$ & $272.3 \pm 8.6$  & $(1.9 \pm 0.44)\times10^{-3}$ \\
                             & Warm & $691.7 \pm 14.6$ & $476.4 \pm 10.2$ & $(1.6 \pm 0.3)\times10^{-5}$ \\
\enddata
\tablecomments{Each row corresponds to a 29+61 day joint fit with a shared $M_{\rm dust}$ per (Type, Component) and epoch-specific temperatures $T_{29}$ and $T_{61}$.}
\end{deluxetable}

\begin{deluxetable*}{c c c c c c c c}
\tablecolumns{8}
\tablewidth{0pt}
\tabletypesize{\footnotesize}
\tablecaption{Elemental mass fractions (percent by mass relative to the total ejecta mass, $M_{\rm ej}=0.059\pm0.011~M_\odot$), recalculated from individual and joint dust masses for each composition.}
\tablehead{
\colhead{Epoch} & \colhead{Component} & \colhead{Composition} &
\colhead{Si (\%)} & \colhead{Mg (\%)} & \colhead{O (\%)} & \colhead{C (\%)} & \colhead{Fe (\%)}
}
\startdata
\multirow{10}{*}{29~days}
  & \multirow{5}{*}{Cold}
    & AmC      & \nodata & \nodata & \nodata & $3.4 \pm 0.8$ & \nodata \\
  & & Fo       & $1.9 \pm 0.7$ & $3.2 \pm 1.2$ & $3.7 \pm 1.4$ & \nodata & \nodata \\
  & & En       & $2.6 \pm 0.9$ & $2.2 \pm 0.8$ & $4.6 \pm 1.6$ & \nodata & \nodata \\
  & & SiO$_2$  & $4.3 \pm 1.5$ & \nodata        & $4.8 \pm 1.7$ & \nodata & \nodata \\
  & & Fe       & \nodata & \nodata & \nodata & \nodata & $2.2 \pm 0.6$ \\
\cline{2-8}
  & \multirow{5}{*}{Warm}
    & AmC      & \nodata & \nodata & \nodata & $0.020 \pm 0.005$ & \nodata \\
  & & Fo       & $0.024 \pm 0.007$ & $0.043 \pm 0.012$ & $0.054 \pm 0.015$ & \nodata & \nodata \\
  & & En       & $0.033 \pm 0.010$ & $0.029 \pm 0.009$ & $0.056 \pm 0.017$ & \nodata & \nodata \\
  & & SiO$_2$  & $0.059 \pm 0.018$ & \nodata           & $0.066 \pm 0.020$ & \nodata & \nodata \\
  & & Fe       & \nodata & \nodata & \nodata & \nodata & $0.014 \pm 0.004$ \\
\hline
\multirow{10}{*}{61~days}
  & \multirow{5}{*}{Cold}
    & AmC      & \nodata & \nodata & \nodata & $2.4 \pm 1.7$ & \nodata \\
  & & Fo       & $3.4 \pm 3.0$ & $6.0 \pm 5.3$ & $7.4 \pm 6.6$ & \nodata & \nodata \\
  & & En       & $4.7 \pm 4.1$ & $3.9 \pm 3.4$ & $8.1 \pm 7.1$ & \nodata & \nodata \\
  & & SiO$_2$  & $7.8 \pm 6.8$ & \nodata       & $8.8 \pm 7.7$ & \nodata & \nodata \\
  & & Fe       & \nodata & \nodata & \nodata & \nodata & $2.5 \pm 1.8$ \\
\cline{2-8}
  & \multirow{5}{*}{Warm}
    & AmC      & \nodata & \nodata & \nodata & $0.00069 \pm 0.00026$ & \nodata \\
  & & Fo       & $0.0054 \pm 0.0014$ & $0.0094 \pm 0.0024$ & $0.012 \pm 0.003$ & \nodata & \nodata \\
  & & En       & $0.0074 \pm 0.0020$ & $0.0064 \pm 0.0017$ & $0.013 \pm 0.003$ & \nodata & \nodata \\
  & & SiO$_2$  & $0.012 \pm 0.003$ & \nodata              & $0.013 \pm 0.004$ & \nodata & \nodata \\
  & & Fe       & \nodata & \nodata & \nodata & \nodata & $0.0010 \pm 0.0003$ \\
\hline
\multirow{10}{*}{Joint~29~+~61~days}
  & \multirow{5}{*}{Cold}
    & AmC      & \nodata & \nodata & \nodata & $4.7 \pm 1.4$ & \nodata \\
  & & Fo       & $2.0 \pm 0.8$ & $3.5 \pm 1.4$ & $2.8 \pm 1.1$ & \nodata & \nodata \\
  & & En       & $2.8 \pm 1.1$ & $2.4 \pm 1.0$ & $4.9 \pm 1.9$ & \nodata & \nodata \\
  & & SiO$_2$  & $4.8 \pm 1.9$ & \nodata       & $5.1 \pm 2.0$ & \nodata & \nodata \\
  & & Fe       & \nodata & \nodata & \nodata & \nodata & $3.2 \pm 0.9$ \\
\cline{2-8}
  & \multirow{5}{*}{Warm}
    & AmC      & \nodata & \nodata & \nodata & $0.042 \pm 0.012$ & \nodata \\
  & & Fo       & $0.014 \pm 0.005$ & $0.025 \pm 0.008$ & $0.033 \pm 0.010$ & \nodata & \nodata \\
  & & En       & $0.019 \pm 0.006$ & $0.017 \pm 0.006$ & $0.029 \pm 0.009$ & \nodata & \nodata \\
  & & SiO$_2$  & $0.033 \pm 0.011$ & \nodata           & $0.037 \pm 0.012$ & \nodata & \nodata \\
  & & Fe       & \nodata & \nodata & \nodata & \nodata & $0.027 \pm 0.008$ \\
\enddata
\tablecomments{ 
For AmC (amorphous carbonaceous) dust, all mass is attributed to C; for silicates (Fo, En, SiO$_2$), the Si, Mg, and O abundances are derived from the stoichiometric ratios of each composition and scaled by the total ejecta mass; and for metallic Fe dust, all mass is attributed to Fe.  
Uncertainties include propagated contributions from both $M_{\rm dust}$ and $M_{\rm ej}$.}
\label{tab:abundances}
\end{deluxetable*}

\section{Abundance Constraints from Kilonova Models}
\label{sec:kilonova_model_comparison}

A prerequisite for the formation of dust within KN ejecta is the formation of alpha-elements (i.e., O, Mg, Si). While typically associated with heavy, $r$-process element production, some lighter elements will be produced in a KN either as a decay-product of heavy nuclei, in regions of particularly high electron fraction ($Y_e = n_p/(n_p+n_n)$), or in an early nucleosynthetic freeze out (e.g., \citealt{Perego2022}).

To explore the possibility of these lighter elements being produced in sufficient quantities to form dust at the level required to produce the thermal emission in GRB 230307A, we utilize nuclear-network calculations from \textit{SkyNet} \citep{Lippuner2017}. We employ KN-like models where each model has a single $Y_e$.  We explore a $Y_{e}$ range from 0.04 (extremely neutron-rich) to 0.49 (roughly equal number of protons and neutrons; typical of SN ejecta) in increments of $\Delta Y_e = 0.05$. Each model assumes typical neutron-star merger ejecta values of $M_{\rm ej} = 0.025\msun$ and $v_{\rm ej} = 0.15c$.  We track abundances out to $t = 100$~days after the merger, covering the phases of interest for \irkilo.

These single $Y_e$ models serve to illustrate the dramatic effect of neutron abundance on the resultant nucleosynthesis, with $r$-process production beyond the first peak ($Z \gtrsim 40$, e.g., Te) vanishing precipitously for $Y_e \gtrsim 0.3$. In reality, the neutron star merger ejecta will not be characterized by a single $Y_e$ and will instead exhibit a  \emph{distribution} representative of the range of physical conditions experienced in different parts of the ejecta (e.g., tidal-tails, accretion-disk winds).

\citet{Anand2023} modeled AT~2017gfo using updated KN models calculated with the 3D Monte Carlo radiative transfer code \textsc{POSSIS}. The models include updated nuclear-heating rates, thermalization efficiencies, $Y_e$- and temperature-dependent opacity, and $Y_e$ distributions informed by merger simulations \citep{Bulla2023}. The models have six parameters describing two ejecta components: a dynamical component with $[M_{\rm dyn}, v_{\rm dyn},  \langle{Y_{e,\text{dyn}}\rangle}]$, where $\langle\rangle$ denotes a mass-average, and a wind component with $[M_{\rm wind}, v_{\rm wind},  {Y_{e,\text{wind}}}]$. 

\begin{figure}
    \centering
    \includegraphics[width=\linewidth]{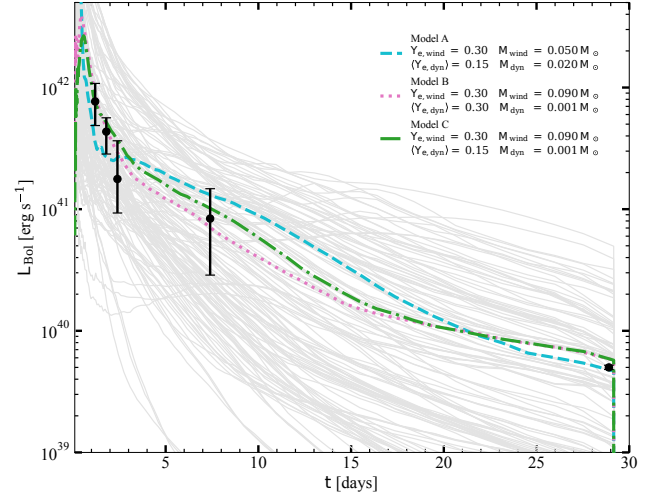}
    \caption{Pseudo-bolometric light curve of \irkilo{} \citep[black points;][]{Liu2025}.  Bolometric light curves of \protect\textsc{POSSIS} KN models are also shown.  Most models are inconsistent with the evolution of \irkilo; the colored curves represent the models with the lowest $\chi_{\nu}^{2}$.  The blue dashed, pink dotted, and green dot-dashed curves represent Models A, B, and C, respectively (see Table~\ref{tab:top3_models}).}
\label{fig:lbol_lcs}
\end{figure}

Figure \ref{fig:lbol_lcs} displays the pseudo-bolometric light curve for AT~2023vfi as calculated by \citet{Liu2025}, along with light curves (in gray) from a coarsely-sampled subset of the \textsc{POSSIS} models described above. We do not perform a full Monte Carlo inference on this model space, but instead highlight (in color) the three best-fitting models according to the reduced chi-squared statistic, again noting that the full model space is undersampled. While all three models have a massive wind-component at $Y_e = 0.3$, there is significant variation in the allowed $Y_e$ distribution for the less massive dynamical ejecta which will result in different heavy-element abundance production. Since these models represent the full qualitative range of $Y_{e}$ distributions, we consider their detailed examination to be representative of all KN models consistent with \irkilo. We list the parameters of the models in Table \ref{tab:top3_models}.

\begin{deluxetable}{ccccc}[ht!]
\tabletypesize{\footnotesize}
\tablewidth{\columnwidth}
\tablecaption{Top Three Kilonova Models Ranked by Reduced $\chi^2$\label{tab:top3_models}}
\tablehead{
\colhead{Model} &
\colhead{$M_{\mathrm{dyn/wind}}$ [$M_\odot$]} &
\colhead{$Y_{e,\langle \mathrm{dyn} \rangle/\mathrm{wind}}$} &
\colhead{$v_{\mathrm{dyn/wind}}$ [$c$]} &
\colhead{$\chi^2_{\mathrm{reduced}}$}
}
\startdata
 A & 0.020/0.050 & 0.15/0.30 & 0.120/0.030 & 1.07 \\
 B & 0.001/0.090 & 0.30/0.30 & 0.250/0.100 & 3.68 \\
 C & 0.001/0.090 & 0.15/0.30 & 0.250/0.100 & 3.97 \\
\enddata
\end{deluxetable}

\begin{figure*}
    \centering
    \includegraphics[width=\linewidth]{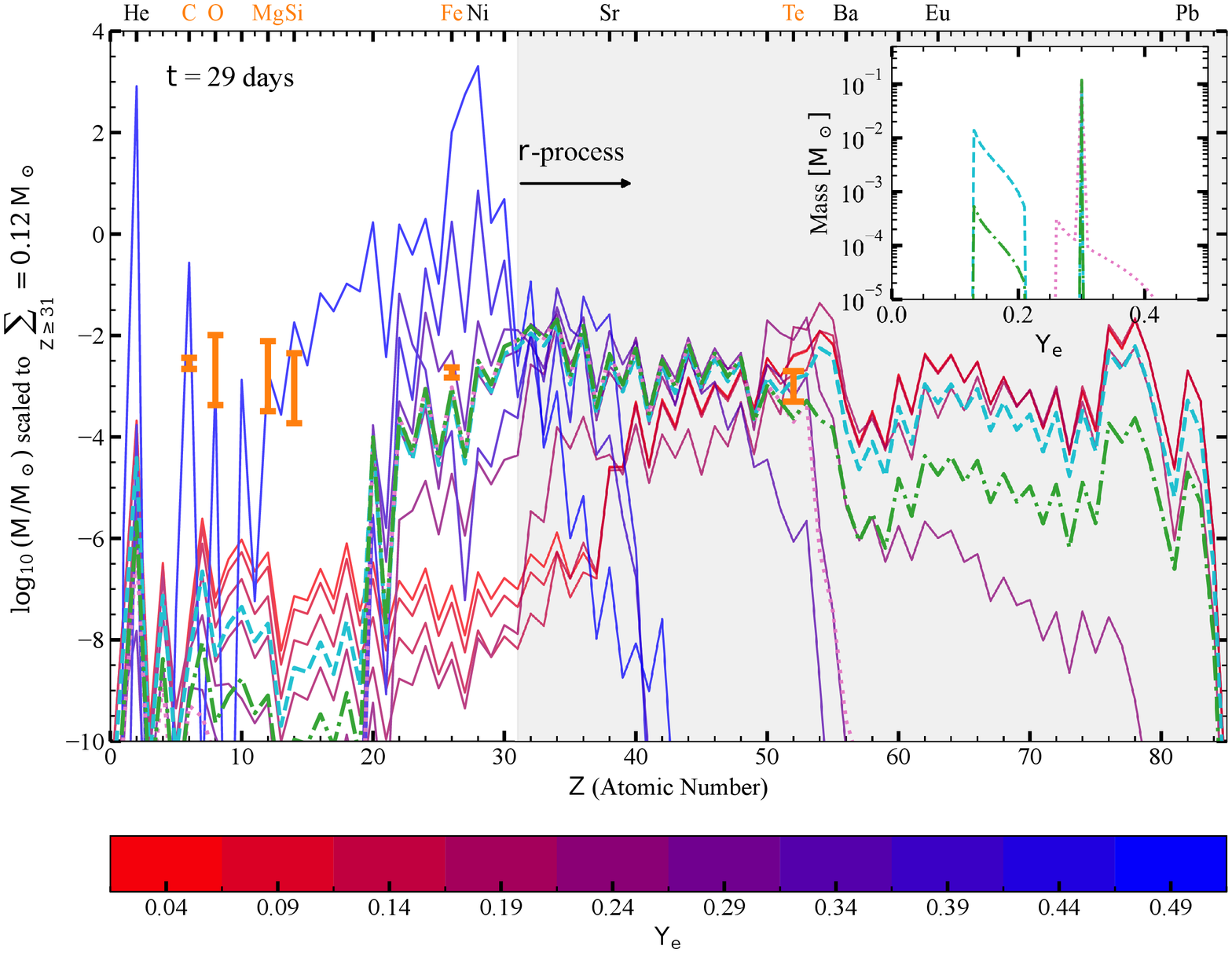}
    \caption{Abundance patterns at $t = 29$~days for single $Y_e$ Skynet runs shown in color, 
    scaled so that the $r$-process mass is equal to $\protect{0.12\,M_\odot}$. 
    The orange vertical bars show the range of possible C, O, Mg, Si, and Fe masses inferred 
    through dust modeling of this epoch for carbonaceous, silicate (Fo),  and pure-Fe dust, as well as the Te constraints from the emission-line modeling of \citet{Levan2024} with an assumed factor of 2 uncertainty in the mass. 
    The three \protect\textsc{POSSIS} KN models from Figure \ref{fig:lbol_lcs} are shown in the dotted, dashed, and dot-dashed lines with their $Y_e$ distributions shown in the inset. 
    No model is able to satisfy the inferred dust and Te masses simultaneously 
    (see Table~\ref*{tab:sigma_devs}).}
    \label{fig:abundances}
\end{figure*}

Figure \ref{fig:abundances} shows the abundances for the single-$Y_e$ \texttt{SkyNet} and the three bolometrically-consistent \textsc{POSSIS} models (see Figure~\ref{fig:lbol_lcs}) at $t = 29$~days post-merger, corresponding to the first \textit{JWST} spectral epoch. To calculate the abundances for the \textsc{POSSIS} models we sum together those from the single $Y_e$ \texttt{SkyNet} models using the appropriate weights informed by the $Y_e$ distributions. We have normalized each model so that the total $r$-process mass (here defined as elements with proton number $Z \geq 31$, or Ga and heavier) is equal to 0.12~\msun, 1.5 times the \textasciitilde0.08~\msun\, inferred by modeling of the pseudo-bolometric light curve of AT~2023vfi \citep{Yang2024,Liu2025}.  We consider this mass to be a conservative upper limit on the total amount of $r$-process material created in GRB~230307A, which sets the mass scale and allows us to calculate absolute masses of alpha-elements (C,O,Mg,Si) to be compared to our silicate and carbonaceous dust constraints, Fe to compare to our pure iron dust constraints, and Te to compare to the $10^{-3}\,M_\odot$ inferred by \citet{Levan2024}, assuming a factor of 2 uncertainty in the mass. For clarity, we only display the silicate-dust constraints assuming a forsterite (Fo, Mg$_2$SiO$_4$) composition as other compositions yield similar constraints to within a factor of $\sim$2. 

\begin{deluxetable}{lcccccc}
\tabletypesize{\footnotesize}
\setlength{\tabcolsep}{4pt}  
\tablecaption{Signed deviations in units of $\sigma$ in log-space, computed assuming a forsterite composition for silicate dust. Entries with $|\Delta_\sigma|\le 3$ are \textit{italicized} to indicate consistency. We assume that the carbonaceous and iron-rich dust is composed purely of elemental C and Fe, respectively, allowing a direct equivalence between dust mass and elemental mass. The Te mass is $10^{-3}\,M_\odot$ \citep{Levan2024} with an assumed factor of 2 for the 1$\sigma$ uncertainty. Model A/B/C are KN yields corresponding to $Y_e$ distributions matching those shown in Fig.~\ref{fig:abundances}. \label{tab:sigma_devs}}
\tablehead{
\colhead{Model} & \colhead{$\Delta_{\sigma}(\mathrm{C})$} & \colhead{$\Delta_{\sigma}(\mathrm{O})$} &
\colhead{$\Delta_{\sigma}(\mathrm{Mg})$} & \colhead{$\Delta_{\sigma}(\mathrm{Si})$} &
\colhead{$\Delta_{\sigma}(\mathrm{Fe})$} & \colhead{$\Delta_{\sigma}(\mathrm{Te})$}
}
\startdata
$Y_e=0.04$ & $-46.67$ & $-6.50$  & $-5.04$  & $-5.94$  & $-42.02$ & $+\textit{1.96}$ \\
$Y_e=0.09$ & $-48.20$ & $-6.79$  & $-5.45$  & $-6.32$  & $-45.38$ & $+\textit{2.07}$ \\
$Y_e=0.14$ & $-48.51$ & $-7.35$  & $-6.29$  & $-7.20$  & $-53.22$ & $+\textit{0.79}$ \\
$Y_e=0.19$ & $-49.46$ & $-8.33$  & $-7.42$  & $-8.70$  & $-56.65$ & $+3.89$ \\
$Y_e=0.24$ & $-51.71$ & $-9.02$  & $-9.33$  & $-10.83$ & $-19.97$ & $+3.04$ \\
$Y_e=0.29$ & $-58.03$ & $-10.46$ & $-12.49$ & $-13.21$ & $-3.68$  & $-\textit{0.73}$ \\
$Y_e=0.34$ & $-109.52$& $-20.70$ & $-22.57$ & $-21.96$ & $+4.25$  & $-10.18$ \\
$Y_e=0.39$ & $-131.25$& $-28.13$ & $-22.04$ & $-17.92$ & $-11.56$ & $-71.64$ \\
$Y_e=0.44$ & $-75.43$ & $-10.89$ & $-14.49$ & $-13.33$ & $+29.72$ & $-88.78$ \\
$Y_e=0.49$ & $+18.03$ & $-\textit{0.47}$ & $+\textit{0.12}$ & $+\textit{1.88}$ & $+47.37$ & $-97.76$ \\
Model A    & $-52.70$ & $-8.15$  & $-7.07$  & $-7.99$  & $-4.37$  & $+\textit{0.47}$ \\
Model B    & $-60.38$ & $-10.07$ & $-9.12$  & $-10.04$ & $-\textit{2.96}$ & $-\textit{2.11}$ \\
Model C    & $-61.36$ & $-11.11$ & $-13.21$ & $-14.10$ & $-\textit{2.90}$ & $-\textit{2.33}$ \\
\enddata
\end{deluxetable}

Table \ref{tab:sigma_devs} quantifies the inability of any single-$Y_e$ model to simultaneously match the Te mass (inferred from modeling the [\ion{Te}{3}] emission line) and light elements or Fe needed to produce sufficient dust to account for the IR continuum. While some low-$Y_e$ models ($Y_e \lesssim 0.25$) are able to match the Te abundance to within 3$\sigma$ when normalized to the inferred $r$-process mass (i.e., they produce the correct mass fraction), they all significantly underproduce any alpha elements by several orders of magnitude compared to the dust constraints. Conversely, our highest-$Y_e$ model plotted is able to roughly satisfy the alpha-element (and thereby dust) constraints, but fails to produce Te in any significant quantity. Additionally, the high $Y_e$ models produce significantly more Fe-group material and much less (by mass) $r$-process, leading our normalization scheme of requiring 0.12~\msun\, of $r$-process ejecta to be accompanied by the production of $\gg$100~M$_{\sun}$ He and Fe-group material. 

The three POSSIS models described above qualitatively span the space of $Y_e$ distributions, as shown in the inset of Figure~\ref{fig:abundances}. Each model includes a massive ($M_{\rm wind}$ = 0.05--0.09~\msun) wind component at $Y_e=0.3$. The dynamical component differs between models, sampling a range of neutron-richness $\langle Y_{e,\text{dyn}} \rangle=0.15,0.3$ and more than an order of magnitude in mass ($M_{\rm dyn}$ = 0.001--0.02~\msun). This is reflected in their resultant heavy-element production, with the high mass, high $Y_e=0.3$ wind producing most of the $20 < Z < 55$ elements, and the low $\langle Y_{e,\text{dyn}} \rangle=0.15$ of Models A and C contributing to $Z > 55$ at their respective dynamical-mass scales. Model B does not produce appreciable mass for $Z > 55$. As seen in Table \ref{tab:sigma_devs}, Models B and C are marginally consistent ($\sim$3~$\sigma$) with both the Fe and Te constraints.

We can also constrain the maximum amount of $^{56}$Ni that could be contributing to the late-time heating by considering it in isolation and  assuming all late-time luminosity is powered by the $^{56}\mathrm{Ni} \rightarrow\, ^{56}\mathrm{Co} \rightarrow\, ^{56}\mathrm{Fe}$
decay chain. This provides an  upper limit to the amount of $^{56}$Ni produced in the ejecta, regardless of the source (neutron-star merger, collapsar, neutron-star white-dwarf merger). We focus only on the JWST epochs (+29 and +61 days) as these are the least affected by possible afterglow contamination and have the tightest constraints on $L_{\rm Bol}$ due to their spectral coverage. For both epochs we use the 3$\sigma$ upper limit on the bolometric luminosity.

The instantaneous heating rate for the 
${}^{56}\mathrm{Ni} \rightarrow {}^{56}\mathrm{Co} \rightarrow {}^{56}\mathrm{Fe}$
decay chain is given by

\begin{equation}
\epsilon(t)
= \epsilon_{\mathrm{Ni}}\, e^{-t/\tau_{\mathrm{Ni}}}
+ \epsilon_{\mathrm{Co}}\, \left( e^{-t/\tau_{\mathrm{Co}}} - e^{-t/\tau_{\mathrm{Ni}}} \right),
\end{equation}

where $\epsilon_{\mathrm{Ni}} = 3.9\times10^{10}~\mathrm{erg\,s^{-1}\,g^{-1}}$, 
$\tau_{\mathrm{Ni}} = 8.8~\mathrm{days}$,
$\epsilon_{\mathrm{Co}} = 6.8\times10^{9}~\mathrm{erg\,s^{-1}\,g^{-1}}$, and 
$\tau_{\mathrm{Co}} = 111.3~\mathrm{days}$.

The thermalization of the input energy is dependent on the unknown density structure, kinematics, and composition of the ejecta, so we bracket the possible thermalization fraction $f_{\rm th}$ assuming two possible functional forms: a simple gamma-ray trapping efficiency $f_{\rm th}(t) = 1 - \exp\left[-(t_0/t)^2\right]$, where we take $t_0 = 35$ days, and a KN-motivated fit given by $f_{\rm th}(t) = 0.36 \left[ \exp(-a t) + \frac{\ln(1 + 2 b t^d)}{2 b t^d} \right]$ \citep{Barnes2016}, where the coefficients $a,b,d$ are taken from their Table 1 corresponding to their maximum ejecta mass of $M_{\rm ej} = 0.05\,\msun$ and velocities $v=0.1,\,0.2,\,0.3\,c$. Using this, the thermalization fraction $f_{\rm th}$ at +29(+61) days is $0.07$–$0.77$ ($0.04$–$0.28$), corresponding to an initial Ni mass of $M_{\rm Ni,0} = 5.60 \times 10^{-4}$–$6.02 \times 10^{-3}$ ($3.76 \times 10^{-4}$–$2.63 \times 10^{-3}$)~\msun. These constraints imply that any model that produces significantly more than $2.6 \times 10^{-3},\msun$ (the upper limit of the more constraining 61-day epoch) of radioactive $^{56}$Ni is ruled out by the data at the 3$\sigma$ level. We note that only our $Y_e=0.49$ model violates this constraint due to the precipitous decline in $^{56}$Ni production below $Y_e=0.5$ in neutron-star merger conditions \citep{Seitenzahl2008}.

\section{Discussion}
\label{sec:discussion}

We have shown that the IR continuum seen in the {\it JWST} spectra of \grb\ can be explained by $\sim${}$3 \times 10^{-3}$, $\sim${}$6 \times 10^{-3}$~M$_{\sun}$, and $\sim${}$2 \times 10^{-3}$~M$_{\sun}$  of carbonaceous, silicate, or metallic dust, respectively.  No POSSIS KN model is able to produce sufficient alpha elements and $r$-process elements to match the [\ion{Te}{3}] and IR continuum, assuming it is dust emission.  Examining $\delta$-function $Y_{e}$ nucleosynthetic models, we also found no $Y_{e}$ can reproduce the necessary abundances.  Furthermore, the stringent limit on the abundance of $^{56}$Ni combined with the alpha element abundances is also inconsistent with all models.

We conclude that if the IR emission is from newly formed dust, \grb is not the result of a BNS or NSBH merger.  Alternatively, if it was caused by a BNS or NSBH merger, then the IR continuum cannot be from newly formed dust.

We examine the physical implications of these results below.

\subsection{Dust as an Explanation for the IR Emission}

Dust formation in KNe is expected to be difficult since the low ejecta mass and high ejecta velocity results in a low number density that makes dust formation difficult and prohibitive for $r$-process dust \citep{Takami2014}.  This becomes harder for more massive elements, which further reduces the number density.

It is possible that rather than forming dust that there was pre-existing dust that was warmed by \grbb.  While this could be a solution to the observations, it is unlikely.

The offset from the host galaxy is $\sim$40~kpc in projection.  There is no detected star formation at or near the GRB location.  It is therefore unlikely that the progenitor system was formed in situ, but rather migrated from a closer galactocentric distance to its position at the time of the GRB.  In this process, it is unlikely that any supernova remnant or significant circumbinary material would be present at the time of the GRB.  Additionally, no H or other lines are detected that may indicate the presence of existing material. \citet{Mereghetti2023} measured a neutral hydrogen density of $N(H) < 5 \times 10^{20}$~cm$^{-2}$, further suggesting a bare environment.

If the dust were pre-existing, the overall interpretation of \irkilo\ would need to be adjusted since the late-time thermal component would be primarily re-processed afterglow emission.  As this scenario is unlikely, we do not explore it further.

A second possibility is that the dust is newly formed and composed of heavy elements.  It is physically unlikely that $r$-process dust formed in any abundance \citep{Takami2014}.  Metallic iron is a possibility, especially since KN ejecta lack the lighter elements that usually bond with iron, reducing the mass of pure iron dust.  Two of our examined KN models (B and C) produce enough Fe and Te to be consistent at the 3-$\sigma$ level with both the dust emission and line emission, respectively.  However, the gas density when the ejecta have cooled to the point when iron dust can condense ($T \approx 800$~K) is too low to efficiently form iron dust \citep[assuming kinematics from \citealt{Yang2024}]{Nozawa2011}.  We therefore find iron dust to be an unlikely possibility.

The alternative is that \irkilo{} is not a KN.  Many transients have significant alpha-element abundance necessary for dust formation; however, few such events would also produce a GRB or synthesize significant $r$-process material. A collapsar can produce a GRB and potentially a significant amount of $r$-process elements. \citep{MacFayden1999, Kohri2005, Siegel2019}.  These explosions also synthesize a large amount of $^{56}$Ni, inconsistent with our constraints.  Dust extinction cannot easily hide $^{56}$Ni heating, as it would necessitate that the heated dust would re-emit at wavelengths redder than 5~\umm, requiring cold dust.  A collapsar at high redshift does not alleviate the tension between the $^{56}$Ni and Te abundance.

Another alternative proposed for \grb{} is a white dwarf-NS (WDNS) merger \citep{Chen2024, Wang2024, Cheong2025, Chrimes2025}.  Although not studied in detail, such a system could expel a significant amount of C and O that could potentially form dust.

WDNS mergers are not expected to produce a significant amount of $r$-process material \citep{Zenati2020, Bobrick2022, Kaltenborn2023, Gottlieb2025}.  The discovery of GRB~211211A and \grb have focused WDNS merger simulations on $r$-process nucleosynthesis.  \citet{Liu2025_2} found that light $r$-process nucleosynthesis can occur, but no elements with mass number $A > 90$, including Te, are produced.  Therefore, if \grb is the result of a WDNS, which could plausibly produce a GRB, KN, and dust, would likely mean that the 2.1~\um line was misidentified.

A further possibility is that \grb{} was caused by the accretion-induced collapse of a WD to a NS \citep[e.g.,][]{Fryer1999, Metzger2008}.  This intriguing possibility can potentially produce a GRB \citep{Metzger2008} and may have low-$Y_{e}$ material that could create a heavy $r$-process KN \citep{Cheong2025}.  Depending on the total mass of the binary system and timescale to collapse, there could also be CO-rich circumstellar material that could form dust.

\subsection{A Neutron-Star Merger as an Explanation for \grb}\label{ss:kn}

Although it is extremely unlikely that \grb{} was the result of a BNS or NSBH merger where the IR continuum is from dust, an alternative is for the continuum to come from an alternative source.

Other than the prominent lines around 2.1 and 4.4~\umm, the IR emission is quite smooth and similar to a blackbody continuum.  Without dust as an emitting source, this continuum must be created by the KN.  The nebular emission of a KN is expected to be composed of several broad forbidden emission lines, and while the lines will blend some, no KN model predicts a smooth continuum, let alone a pseudo-blackbody continuum \citep[e.g.,][]{Hotokezaka2021, Hotokezaka2022, Banerjee2025, Pognan2025}.

If the KN is optically thick, there should be a pseudo-blackbody continuum.  However in that case, there should also be P-Cygni profiles from strong permitted line transitions \citep{Watson2019, Domoto2022, Sneppen2023, Tarumi2023}, which are not seen in \irkilo.  Physically, this is an unlikely scenario since it would require especially high densities at even +61~days, while some KN models become optically thin within a week of merger \citep[e.g.,][]{Kasen2019}. Furthermore, forbidden fine-structure transitions such as [\ion{Te}{3}] require low electron densities ($n_e \lesssim 10^{6}$–$10^{7}$~cm$^{-3}$) so that radiative de-excitation dominates over collisional quenching \citep{Hotokezaka2023, Mulholland2024}. These conditions are expected only once the emitting region becomes optically thin. Sustaining a photospheric continuum while producing strong [\ion{Te}{3}] emission would therefore demand simultaneously high and low densities across a broad velocity range, inconsistent with a standard KN.

At these late times, models predict emission from several forbidden transitions of heavy elements \citep{Hotokezaka2021, Hotokezaka2022,Hotokezaka2023, Tanaka2023, Banerjee2025,McCann2025}.  The high ejecta velocity and density of line transitions cause lines to overlap, often creating a pseudo-continuum.  However, no model has such a smooth continuum as seen in \irkilo.

\section{Conclusions}
\label{sec:conclusions}

\grb is an unusually bright, long-duration gamma-ray burst whose faint afterglow and offset from its putative host galaxy resemble those of short GRBs associated with compact-object mergers \citep{Levan2024, Yang2024}. Critically, it is the first GRB for which deep late-time \textit{JWST} IR spectra have been obtained \citep{Levan2024, Gill&Smartt2025}. These spectra exhibit a smooth, blackbody-like thermal continuum, providing a compelling motivation to test whether the observed IR excess can be explained by emission from newly formed dust in the ejecta.

We have tested this possibility by modeling the +29- and +61-day \textit{JWST}/NIRSpec spectra with physically motivated dust emission models. Using an MCMC framework, we have decomposed the spectra into afterglow, emission lines, and a thermal continuum modeled as a modified blackbody emission from carbonaceous, silicate, or metallic iron dust. We fit two-component dust models to match the continuum shape at both epochs, yielding total dust masses of $\sim${}$3 \times 10^{-3}$~M$_{\sun}$ for amorphous carbon, $\sim${}$6.0 \times 10^{-3}$~M$_{\sun}$ for silicates, and $\sim${}$2.0 \times 10^{-3}$~M$_{\sun}$ for iron dust.

We find that no combination of KN nucleosynthetic models, either from \textsc{POSSIS} or from single–$Y_e$ \texttt{SkyNet} calculations, can easily account for the inferred dust-forming elements and the observed [\ion{Te}{3}] emission. The required abundances of C, O, Mg, and Si are orders of magnitude higher than expected from neutron-star merger ejecta.  Two of our examined KN models are mildly consistent (at the 3-$\sigma$ level) with the Fe and Te mass required for metallic iron dust and the [\ion{Te}{3}] emission, but but producing iron dust in \irkilo{} is implausible.

If the IR excess indeed arises from newly formed dust, the nucleosynthetic inconsistencies effectively rule out a BNS or NSBH merger origin for \grbb. Pre-existing dust heated by the afterglow is also disfavored given the large projected offset and low number densities. Dust formation from heavy $r$-process species or metallic iron is physically implausible under the expected ejecta densities and temperatures. Alternative progenitor scenarios—such as WDNS mergers or accretion-induced collapse events—may better reconcile the production of C- and O-rich ejecta with the observed dust emission, although these channels are not expected to synthesize Te in measurable amounts.

Any model for \grb must account for both the IR continuum and emission features.  No known model appears to satisfy this criterion.

When the next similar event is discovered, more late-time epochs and data extending further into the IR would better track the evolution of the continuum and line features.  With these data, one could see if the the data are consistent with dust growth, which would better constrain any dust model.

\begin{acknowledgments}
We thank S.E.\ Woosley for extended discussions about nucleosynthesis. P.A.\ thanks Kirsty Taggart for valuable discussions on the dust modeling and Kaew Tinyanont for the python script \texttt{sed\_et\_al} used in the dust model analysis. P.M.\ thanks Luke Roberts for providing the \texttt{SkyNet} data and Mattia Bulla for making the \textsc{POSSIS} models publicly available.

The UCSC transients team is supported in part by NSF grant AST--2307710 and by a fellowship from the David and Lucile Packard Foundation to R.J.F.

This work is based in part on observations made with the NASA/ESA/CSA James Webb Space Telescope, which is operated by the Association of Universities for Research in Astronomy, Inc., under NASA contract NAS 5-03127 for JWST. These observations are associated with programs GO--4434 and GO--4445 (PI Levan).

This work was partially performed at the Aspen Center for Physics, which is supported by National Science Foundation grant PHY--2210452.

\end{acknowledgments}

\software{\texttt{numpy} \citep{harris2020array}, \texttt{matplotlib} \citep{Hunter:2007}, \texttt{astropy} \citep{2013A&A...558A..33A,2018AJ....156..123A,2022ApJ...935..167A}, \texttt{emcee} \citep{Foreman-Mackey_etal_2013_emcee}}

\bibliographystyle{aasjournal}
\bibliography{ref.bib}
\end{document}